\documentclass[10pt,a4paper,DIV8]{scrartcl} 
\usepackage[utf8x]{inputenc}
\usepackage[T1]{fontenc}
\usepackage{lmodern}
\usepackage[british]{babel}
\usepackage{geometry}
\usepackage{float}
\usepackage{textcomp}
\usepackage{amsmath}
\usepackage{amssymb}
\usepackage{amsfonts}
\usepackage{graphicx} 
\usepackage{pgf}
\usepackage{xcolor}
\usepackage{hyperref}
\usepackage{authblk}
\usepackage{bbold} 

\newcommand*\samethanks[1][\value{footnote}]{\footnotemark[#1]}
\newcommand{\arxiv}[1]{arXiv:\,\href{http://arxiv.org/abs/#1}{\texttt{#1}}}
\newcommand{\Tr}{\ensuremath{\mathrm{Tr}}}
\newcommand{\aetap}{\ensuremath{\text{a-}\eta'}}
\newcommand{\afn}{\ensuremath{\text{a-}f_0}}

\makeatletter
\usepackage{bbm}
\usepackage{slashed}
\usepackage{braket}
\usepackage{todonotes}
\makeatother
\title{%
   {\vspace{-20mm}\normalsize
    \hfill\parbox[b][30mm][t]{35mm}{\textmd{MS-TP-19-01}}}\\[-18mm]
Variational analysis of low-lying states in supersymmetric Yang-Mills theory
\vspace*{2mm}}
\author[1,2]{Sajid Ali%
\thanks{\{sajid.ali,h.gerber,simon.kuberski,munsteg,scior\}@uni-muenster.de}}
\author[3,1]{Georg~Bergner\thanks{georg.bergner@uni-jena.de}} 
\author[1]{Henning~Gerber\samethanks[1]}
\author[1]{Simon~Kuberski\samethanks[1]}
\author[4]{Istvan~Montvay\thanks{montvay@mail.desy.de}}
\author[1]{Gernot~M\"unster\samethanks[1]}
\author[5]{Stefano~Piemonte\thanks{stefano.piemonte@ur.de}}
\author[6]{Philipp~Scior\samethanks[1]}
\affil[1]{University of M\"unster, Institute for Theoretical Physics,
Wilhelm-Klemm-Str.~9, D-48149 M\"unster, Germany}
\affil[2]{Government College University Lahore, Department of Physics,
Lahore 54000, Pakistan}
\affil[3]{University of Jena, Institute for Theoretical Physics,
Max-Wien-Platz 1, D-07743 Jena, Germany}
\affil[4]{Deutsches Elektronen-Synchrotron DESY,
Notkestr.~85, D-22607 Hamburg, Germany}
\affil[5]{University of Regensburg, Institute for Theoretical Physics,
Universit\"atsstr.~31, D-93040 Regensburg, Germany}
\affil[6]{Universit\"at Bielefeld, Fakult\"at f\"ur Physik,
Universit\"atsstr.~25, D-33615 Bielefeld, Germany}
\date{January 08, 2019} 
\begin{document}

\maketitle

\newpage

\begin{abstract}
\noindent
\textbf{\textsf{Abstract:}}
We have calculated the masses of bound states numerically in $\mathcal{N}=1$
supersymmetric Yang-Mills theory with gauge group SU(2). Using the suitably
optimised variational method with an operator basis consisting of smeared
Wilson loops and mesonic operators, we are able to obtain the masses of the
ground states and first excited states in the scalar, pseudoscalar and
spin-\textonehalf~sectors. Extrapolated to the continuum limit, the
corresponding particles appear to be approximately mass degenerate and to
fit into the predicted chiral supermultiplets. The extended operator basis
including both glueball-like and mesonic operators leads to improved results
compared to earlier studies, and moreover allows us to investigate the
mixing content of the physical states, which we compare to predictions in
the literature.

\end{abstract}
\section{Introduction}  
 
The $\mathcal{N}=1$ supersymmetric Yang-Mills (SYM) theory is the minimal
supersymmetric extension of the pure gauge sector of QCD, describing the
strong interactions of gluons and their fermionic superpartners, the
gluinos. Unbroken supersymmetry requires the physical bound states of
$\mathcal{N}=1$ SYM to form supermultiplets of degenerate masses. Based on
supersymmetric effective actions, the lowest-lying supermultiplet has
originally been proposed in Ref.~\cite{Veneziano:1982ah} to be a chiral
supermultiplet formed out of a scalar and a pseudoscalar meson and a
spin-\textonehalf~bound state of gluons and gluinos, called gluino-glue. The
original analysis has been extended in
Ref.~\cite{Farrar:1997fn,Farrar:1998rm} to include also glueball operators
which can mix with the meson operators carrying the same quantum numbers.
The scalar and pseudoscalar bound states are therefore created by linear
combinations of meson-like and glueball-like operators. The mass hierarchy
and the amount of mixing are difficult to predict theoretically from the
effective Lagrangian and are in general unconstrained by supersymmetry.

Numerical simulations of $\mathcal{N}=1$ SYM on a space-time lattice are
required to verify and extend the analytical predictions about the low
energy effective theory. The study of the lightest bound states of
$\mathcal{N}=1$ SYM with SU(2) gauge symmetry has been the main subject of
several publications by our collaboration
\cite{Demmouche:2010sf,Bergner:2015adz,Bergner:2013nwa,Bergner:2012rv,Bergner:2014ska}.
The results reveal the expected degeneracy of the members of the chiral
multiplet in the continuum limit. In particular, the scalar glueball and the
scalar meson masses, extrapolated to the continuum limit, have compatible
values, which hints at mixing of the states in this channel.
 
In the present work we have optimised our techniques to extract the masses
with respect to our previous work. This allows for the first time to
investigate the mixing properties of the two lowest supermultiplets from the
lattice gauge ensembles we have generated. In this article we explain these
methods and their optimisations. We show, in particular, that the
construction of an enlarged correlation matrix including both, the meson and
glueball operators, is crucial to extract the physical states in the scalar
channel. Using these techniques we improve the precision of our previous
results for the lightest masses and we are able to extract the masses of the
next heavier states in the scalar, pseudoscalar, and fermion channel, too.
In this way we provide first results concerning the possible formation of a
supermultiplet of excited states on the lattice.

The formation of supermultiplets at the level of the excited states is strong
evidence for the fact that the unavoidable breaking of
supersymmetry on the lattice can be kept under control. The states with
higher masses are affected stronger by lattice artefacts, and hence the
predicted degeneracy of these states is more sensitive to the breaking of
supersymmetry on the lattice. The present work represents the first step
towards an investigation of the spectrum of excited states of
$\mathcal{N}=1$ SYM. An alternative determination of supersymmetry
violations beyond the ground state level would be the weighted
sum of all energy differences in terms of the derivative of the Witten
index, see \cite{Bergner:2015cqa}.

In addition to our results on excited states, we confirm our earlier results
concerning the mixing of glueball and meson operators in the ground state,
and we are able to determine the glueball and the meson contents. Eventually
this might lead to a better understanding of the conjectures concerning the
low energy multiplets of the theory.

In this work we have optimised our methods for SYM with gauge group SU(2).
The techniques are, however, also suitable for other theories. Currently we
are applying the same methods successfully to SYM with gauge group SU(3),
for which the results will be published soon. Our techniques are also
applicable to lattice QCD, where similar investigations are being done.

It should be noted that our studies, and corresponding investigations in QCD
as well, are quite challenging due to the noisy signals from glueball
operators and the disconnected meson contributions.

This paper is organised as follows: $\mathcal{N}=1$ SYM in the continuum and
on the lattice is introduced in the next section. The current paper is
particularly focused on the technical aspects, which are explained in
Sections \ref{sec:ComputationOfMasses}, \ref{sec:OptimizingJacobi}, and
\ref{sec:operatorbasis}. The physical results, the improved signal for the
ground states, the mixing, and the excited states, are finally presented in
Sections \ref{sec:multiplet} and \ref{sec:mixing}.

\section{Supersymmetric Yang-Mills theory on the lattice}

The Lagrangian $\mathcal{L}$ of the $\mathcal{N}=1$ SYM in Euclidean
space-time,
\begin{equation}
\mathcal{L} =   
\frac{1}{4} F_{\mu\nu}^{a} F_{\mu\nu}^{a} +
\frac{1}{2} \bar{\lambda}^{a} \gamma_{\mu} (\mathcal{D}_{\mu}
\lambda)^{a}
+ \frac{m_{\text{g}}}{2} \bar{\lambda}^{a} \lambda^{a}\,,
\end{equation}
is similar to the one of one-flavour QCD. The main difference is that
supersymmetry requires the gluino field $\lambda$ to be a Majorana fermion
field in the adjoint representation of the gauge group. Correspondingly the
gauge covariant derivative is the adjoint one, given by $(\mathcal{D}_{\mu}
\lambda)^{a} = \partial_{\mu} \lambda^{a} + g\,f_{abc} A^{b}_{\mu}
\lambda^{c}$. The gluino mass term breaks supersymmetry softly and the
renormalised gluino mass must be tuned to zero to recover the full
supersymmetry. In this work we focus on the theory with gauge group SU(2).
The light mesons in the particle spectrum of SYM are similar to the mesons
in QCD, apart from the difference that the constituent fermions are gluinos
and not quarks. Due to this similarity, the scalar meson is called $\afn$
and the pseudoscalar meson is called $\aetap$ where the ``a'' indicates the
adjoint representation. The mesons are expected to mix with the
corresponding glueballs with the same quantum numbers. In addition to these
particles, the low-energy spectrum contains a spin-\textonehalf~bound state
of gluons and gluinos, called gluino-glue, which has no analogy in QCD.

On the lattice, the fermion part of the action is discretised using the
Wilson-Dirac operator $D_W$, which effectively removes the doublers from the
physical spectrum, but at the same time explicitly breaks chiral symmetry.
As shown in Ref.~\cite{Curci:1986sm}, chiral symmetry and supersymmetry can
be recovered simultaneously in the continuum limit by tuning of the bare
gluino mass. This tuning can be achieved in different ways. One way is to
determine the point where the renormalised gluino mass vanishes from the
supersymmetric Ward identities~\cite{Ali:2018fbq}.
Another way is to employ the adjoint pion
mass, which is defined in a partially quenched approach
\cite{Munster:2014cja}, and to extrapolate to the point where it vanishes.
For non-zero lattice spacings, lattice artefacts introduce a small mismatch
between these determinations. Because the adjoint pion mass can be measured
quite precisely and effectively, we use this quantity to define our
extrapolations to the chiral point. This corresponds to the point where
supersymmetry should be restored in the continuum limit.

$\mathcal{N}=1$ SYM is asymptotically free, and the running of the strong
coupling has been investigated non-perturbatively in
Ref.~\cite{Bergner:2017ytp}. The extrapolation to zero lattice spacing $a
\rightarrow 0$ corresponds to the weak coupling limit $g \rightarrow 0$. The
lattice spacing $a$ used for the extrapolation to the continuum limit is
measured in terms of the infrared scale $w_0$ defined from the gradient
flow~\cite{Luscher:2010iy,Borsanyi:2012zs,Bergner:2014ska}. In order to
suppress lattice discretisation effects, we use a tree level Symanzik
improved action together with stout smearing on the gauge links in the
Wilson-Dirac operator, see~\cite{Bergner:2013nwa} for further details.

The production of the gauge field configurations is performed with the
two-step polynomial hybrid Monte Carlo (PHMC) algorithm
\cite{Montvay:2005tj,Demmouche:2010sf}. A mild sign problem of our lattice
formulation arises from the integration of the Majorana fermions
\cite{Bergner:2011zp} and is corrected by reweighting.

\section{Techniques for the determination of light states}
\label{sec:ComputationOfMasses}

In order to gain insights into the structure of the low energy spectrum of
SYM, consisting of gluino-glue particles as well as mixtures of mesons and
glueballs, we determine their masses as well as the glueball and meson
content in the scalar and the pseudoscalar states. These quantities can be
reliably extracted from the gauge field ensembles using the well-known
variational method for which we have systematically optimised the required
techniques and parameters.

Our approach is to use a set of basic interpolators of the physical states
to which we then apply smearing techniques to create the operator basis used
in the variational method. In Sections \ref{sec:ComputationOfMasses} to
\ref{sec:operatorbasis} we explain in detail the techniques and the tunings
used within this approach.

The mesons in SYM are all flavour-singlet ones, which require techniques for
the measurement of all-to-all propagators. We have found that a
preconditioned stochastic estimation is optimal for our lattice volumes.
This is explained in Sec.~\ref{sec:set}.

\subsection{Computation of the masses using the variational method}

A set of interpolating operators $O_i$ must be chosen in order to extract
information about physical states within the variational method. In
principle the only requirement is that the operators $O_i$ have the same
quantum numbers under the lattice symmetry group
as the physical state of interest. In practice the choice of
this variational basis has a crucial influence on the precision of the
results, and therefore we explain in detail the optimisation of these
operators. To extract the masses of physical states using the variational
method, the correlation matrix is build from the time-slice correlators of
the set $\{O_i\}$
\begin{align}
C_{ij}(t) & =\left\langle O_{i}(t)O_{j}^{\dagger}(0)\right\rangle.
\end{align}
The solution of the generalised eigenvalue problem (GEVP) associated with
the correlation matrix $C(t)$,
\begin{equation}
C(t)\vec{v}^{\,(n)}=\lambda^{(n)}(t,t_{0})C(t_{0})\vec{v}^{\,(n)},
\label{eq:GEVP}
\end{equation}
provides generalised eigenvalues $\lambda^{(n)}$ and their corresponding
eigenvectors $\vec{v}^{\,(n)}$. In Ref.~\cite{Luscher:1990ck} it has been
shown that the generalised eigenvalues $\lambda^{(n)}(t,t_{0})$ satisfy
\begin{equation}
\underset{t\rightarrow\infty}{\lim} \lambda^{(n)}(t,t_{0}) \propto
\mathrm{e}^{-m_{n}\left(t-t_{0}\right)}
\left(1+\mathcal{O}\left(\mathrm{e}^{-\Delta m_{n}(t-t_{0})}\right)\right),
\label{e:lambda_GEVP}
\end{equation}
with
\begin{equation}
\Delta m_{n}=\underset{l\ne n}{\min}\left|m_{l}-m_{n}\right|,
\end{equation}
where $m_{n}$ are the masses of the physical states. In order to supress
contributions of higher excitations, it has been suggested in
Ref.~\cite{Blossier:2009kd} to use $t_0\geq t/2$. This, however, leads to
very noisy eigenvalues $\lambda^{(n)}$ for the correlators considered here
and we therefore use $t_0=0$ or $t_0=1$ when necessary. A first estimate can
be obtained from the effective mass
\begin{align} 
\label{eq:effective mass}
m^{(n)}_{\text{eff}}(t,t_0)
=\ln\frac{\lambda^{(n)}(t,t_0)}{\lambda^{(n)}(t+1,t_0)}\,.
\end{align}
A more precise mass estimation is obtained by fitting the eigenvalues
$\lambda^{(n)}(t,t_{0})$ to an exponential function of $t$.

The set of interpolating fields ideally should have large overlaps with the
physical states of interest and small overlap with higher excited states,
otherwise terms coming from higher order excitations might create large
corrections to the expected exponential behaviour of the eigenvalues.

\subsection{Interpolating operators} \label{sec:Interp.Operators}

\begin{figure}
\begin{center}
\includegraphics[width=0.2\textwidth]{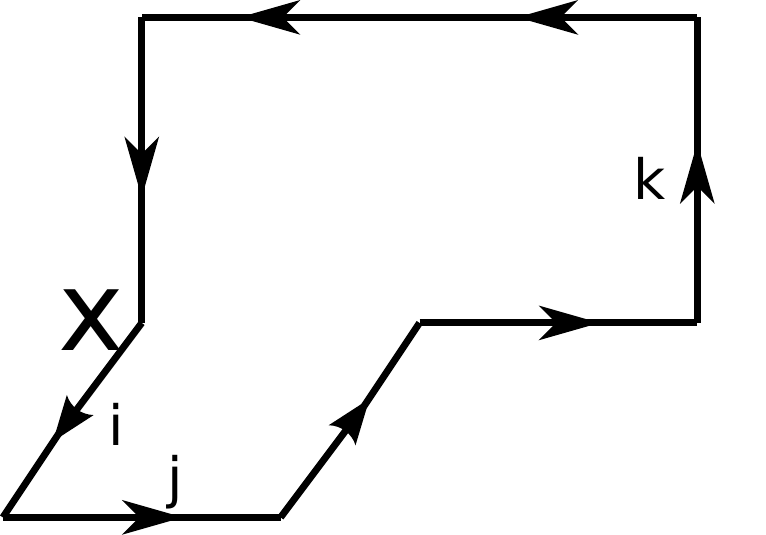}
\protect\caption{Gauge-loop
$\mbox{\ensuremath{\mathcal{C}}}_{ijk}(x)$ used in the interpolating field
of $0^{-+}$-glueball.}
\label{fig:gauge-loops} 
\end{center}
\end{figure}

The basic interpolators for glueballs are built from gauge link loops that
represent the spin and parity quantum number of the respective state. For
the scalar glueball we use a sum of gauge plaquettes
\begin{equation}
O_{\text{gb}^{++}}(x)=\Tr \left[P_{12}(x)+P_{23}(x)+P_{31}(x)\right],
\end{equation}
where $P_{ij}$ denotes a plaquette in the $i$-$j$ plane. For the
pseudoscalar glueball we use
\begin{equation}
O_{\text{gb}^{-+}}(x)
=\sum_{R\in\mathbf{O}_{h}}[\Tr\left[
\mathcal{C}(x)\right] - \Tr\left[P\mathcal{C}(x)\right],
\end{equation}
where the sum is over all rotations of the cubic group $\mathbf{O}_h$, and
$P\mathcal{C}$ is the parity conjugate of the loop $\mathcal{C}$, which is
depicted in Fig.~\ref{fig:gauge-loops}.

The basic interpolating fields for the scalar mesons are
\begin{align}
O_{\text{a-f}_{0}}(x)=&\bar{\lambda}(x)\lambda(x), \qquad 
O_{\aetap}(x)=\bar{\lambda}(x)\gamma_{5}\lambda(x).
\end{align}
When inserted into the correlation matrix, Wick contractions of these fields
lead to connected and disconnected pieces
\begin{multline} 
\label{eq:CorrelationFunction}
\langle\bar{\lambda}(x)\Gamma\lambda(x)\bar{\lambda}(y)\Gamma\lambda(y)\rangle\\
=\Tr\left[\Gamma D_W^{-1}(x,x)\right]\Tr\left[\Gamma D_W^{-1}(y,y)\right]
-2\,\Tr\left[\Gamma D_W^{-1}(x,y)\Gamma D_W^{-1}(y,x)\right],
\end{multline}
where $D_W^{-1}(x,y)$ denotes the propagator from $x$ to $y$ (spin and group
indices suppressed) and $\Gamma$ represents $\mathbb{1}$ or $\gamma_5$.

The interpolating field of the gluino-glue state is given by 
\begin{equation}
 O_{\text{gg}}^{\alpha}(x)
=\sum_{i<j=1}^3\sigma_{ij}^{\alpha\beta}
\Tr\left[P_{ij}(x)\lambda^{\beta}(x)\right]
\quad \text{with} \quad  
\sigma_{\mu\nu} = \frac{\mathrm{i}}{2}[\gamma_\mu,\gamma_\nu].
\label{e:O_gg}
\end{equation}
The gluino-glue correlation matrix consists of an odd and an even part under
time reversal
\begin{equation}
C^{\alpha \beta}_\text{gg} (t)
= C_1(t) \delta^{\alpha \beta} + C_{\gamma_4}(t)\gamma_{4}^{\alpha\beta}.
\end{equation}
Projections to the odd part $C_1(t)$ and to the even part $C_{\gamma_4}(t)$
both provide valid correlators to be used in the GEVP. For the tuning of our
methods we use $C_1(t)$ and for the final results we use a weighted sum of
the results from both $C_1(t)$ and $C_{\gamma_4}(t)$.

\subsection{Stochastic estimator technique (SET)}
\label{sec:set}

The calculation of correlators including more than one gluino field in the 
interpolating operators requires the estimation of the fermion propagator
from all to all lattice points.
Since a complete determination of the inverse of the Dirac-Wilson operator
$D_W$ is prohibitively expensive, we approximate $D_W^{-1}$ by using the
stochastic estimator technique (SET) \cite{Dong:1993pk}, which is improved
by a truncated eigenmode approximation and even-odd preconditioning,
see~\cite{Bergner:2011zz} for further details. The preconditioned
approximation converges faster to the exact result and the numerical
computation of the eigenspace of the preconditioned matrix is much faster
than the one of the full matrix. For the inversion of the preconditioned
Dirac matrix $D_p$ we use its Hermitean version, obtained by multiplication
with $\gamma_5$, since the singular value decomposition provides a better
approximation.

The idea of SET is to solve the Dirac equation on a set of source noise
vectors $\left|\eta^{i}\right\rangle $ fulfilling the relation
\begin{align} \
\frac{1}{N_{S}} \sum_{i}^{N_{S}} \left|\eta^{i}\right\rangle 
\left\langle \eta^{i}\right| 
& =\mathbbm{1}+\mathcal{O}\left(1/\sqrt{N_{S}}\right).
\end{align}
The entries of the vectors $\left|\eta^{i}\right\rangle$
are $\mathbb{Z}_{4}$ complex numbers of
the form $(\pm 1 \pm \textrm{i})/\sqrt{2}$. The propagator $D_p^{-1}$ is
then approximated by
\begin{align}
D_p^{-1} & =\frac{1}{N_{S}} \sum_{i}^{N_{S}} \left|s^{i}\right\rangle 
\left\langle \eta^{i}\right|+\mathcal{O}\left(1/\sqrt{N_{S}}\right),
\quad \text{with} 
\left|s^{i}\right\rangle =D_p^{-1}\left|\eta^{i}\right\rangle,
\label{eq:SET}
\end{align}
and $D_p^{-1}\left|\eta^{i}\right\rangle$ is calculated using a conjugate
gradient solver.

For the truncated eigenmode approximation the $N_E$ lowest eigenvalues
$\lambda_i$ and eigenvectors $|v_i\rangle$ of the Hermitean operator
$\gamma_5 D_p$ are computed. The truncated eigenvector approximation of the
inverse matrix is
\begin{equation}
\label{eq:genapprox}
 D_p^{-1}\approx \sum_{i=1}^{N_E} \frac{1}{\lambda_i} |v_i \rangle \langle v_i |\, .
\end{equation}

The two approximations are easily combined: the noise vectors of SET are
just projected to the subspace orthogonal to the one spanned by the lowest
eigenvectors. In the end both contributions are summed, gaining also a
speedup of the inversions due to the better condition number of the
projected operator.
\begin{equation} \label{eq:SETplusEigenmodes}
D_p^{-1}
\approx \sum_{i=1}^{N_E} \frac{1}{\lambda_i} |v_i \rangle \langle v_i |\, 
+ \frac{1}{N_{S}} \sum_{i}^{N_{S}}
\left|s_\perp ^{i}\right\rangle \left\langle \eta_\perp^{i}\right| \doteq 
\sum_{i=1}^{N_E + N_S} a_i |w_i \rangle \langle u_i |\,.
\end{equation}

For the case of plain Wilson fermions, investigated in the current study,
the even-even part $M_{ee}$ of the Wilson-Dirac matrix is just the identity.
More generally, the inverse of the block diagonal matrix $M_{ee}$ can be
computed exactly. However, in order to simplify the smearing procedure, we
have used a large number of stochastic sources to approximate the inverse of
this matrix. These can be computed very efficiently.

We have optimised the parameters of this approximation in order to reduce
the noise and to speed up the computations. The tuning of the number of noise
vectors required for a reliable estimation of the disconnected piece is
discussed in Sec.~\ref{sec:SETOptimization}. The eigenspace of the
preconditioned matrix computed in the measurement of disconnected
contributions is also used for a deflation of the inversions in other
measurements, like the connected meson or the gluino-glue correlators.
Concerning speedup, the optimisation is also machine dependent, e.~g.\ our
runs on KNL based machines require quite different parameters.

\subsection{Smearing techniques}

The different interpolating fields used in the variational method are
constructed by applying smearing techniques to the basic interpolating
fields defined in \ref{sec:Interp.Operators}.

We use APE-smearing \cite{Albanese:1987ds} on the gauge links in order to
create smeared gluino-glue and smeared glueball operators. We use a smearing
parameter $\epsilon_{\text{APE}}=0.4$ for smearing the gluino-glue
interpolators and $\epsilon_{\text{APE}}=0.5$ for smearing the glueball
interpolators.

For the construction of the fermion source we use gauge invariant Jacobi 
smearing \cite{Gusken:1989ad} \\
\mbox{$\lambda\rightarrow\ \tilde \lambda = F\lambda$} with the smearing operator
\begin{equation}
\label{eq:JacobiSmearing}
F_{\beta b,\alpha a}(\vec{x},\vec{y})
=C_{J}^{N_{J}} \delta_{\beta\alpha}
\left(\delta_{\vec{x},\vec{y}}
+ \sum_{i=1}^{N_{J}}
(H^{i})_{b a}(\vec{x},\vec{y})
\right),
\end{equation}
with
\begin{equation} \label{eq:SmearingKernel}
H_{b a}(\vec{x},\vec{y})
= \kappa_{J}\sum_{i=1}^{3}\left[
\delta_{\vec{y},\vec{x}+\hat{i}}\tilde{U}_{i,ba}(\vec{x})
+\delta_{\vec{y},\vec{x}-\hat{i}}
\tilde{U}_{\hat{i},ba}^{\dagger}(\vec{x}-\hat{i})\right].
\end{equation}
Here $N_{J}$ is the integer Jacobi smearing level, $\kappa_{J}$ is a Jacobi
smearing coefficient, $C_{J}$ is a normalisation constant and
$\tilde{U}_i(x)$ are APE smeared gauge links. The tuning of the parameters
$N_J$, $N_J$ and $C_J$ and the smearing of the gauge links is explained in
Sec.~\ref{sec:OptimizingJacobi}.

Smearing the fermions fields is equivalent to replacing the propagator
$D_W^{-1}$ with the smeared propagator
\begin{align}
D_W^{-1} \rightarrow \tilde{D}_W^{-1}&=FD_W^{-1}F^\dagger\,.
\end{align} 
For the disconnected piece, this translates to smearing the source and sink
vectors in equation (\ref{eq:SETplusEigenmodes})
\begin{equation}
\Tr \left[ \Delta \right] 
\approx\frac{1}{N_{S}} \Tr \left[ \sum_{i,\vec{x}}^{N_{S}} a_i
F\ket{w^{i}}\bra{u^{i}}F^\dagger \right]. 
\label{eq:smearedSET}
\end{equation}
The connected piece is calculated using standard delta sources on which
$F^\dagger$, $D_W^{-1}$ and $F$ are subsequently applied:
\begin{equation} 
\tilde{D}_W^{-1}\delta=FD_W^{-1}  F^\dagger\delta. 
\label{eq:deltaSources}
\end{equation} 

\section{Optimising the methods}  
\label{sec:OptimizingJacobi}

\subsection{Jacobi smearing}

Aiming to achieve a good signal-to-noise ratio for the meson measurements,
we systematically optimise the Jacobi smearing parameters $K_J$, $C_J$, the
smearing level $N_\text{APE}$ of the smeared gauge fields $\tilde{U}$ and
the Jacobi smearing levels $N_J$.
 
As explained in \cite{Ali:2017iof} there is a critical value of the parameter
$\kappa_J^c$. For values smaller than $\kappa_J^c$, the smearing operator
$F$ converges in the limit $N_J\rightarrow \infty$, while for values larger
than $\kappa_J^c$ it diverges. In order to smear efficiently and at the same
time to avoid large numerical errors we choose a value 
\begin{equation}
\kappa_{J}=0.2\,,
\end{equation} 
which is just above $\kappa_J^c$. With this choice the smearing radius is
less than 6 lattices spacings up to smearing levels $N_J=100$
\cite{Ali:2017iof}. The normalisation factor $C_{J}$ is chosen such that the
values of the correlation functions stay at the same order of magnitude for
large and small smearing levels. We use
\begin{equation} 
C_{J}=0.87.
\end{equation}

In principle, the final results do not depend on the values for $\kappa_J$
and $C_J$. The efficiency could, however, depend on the choice of the values
of the smearing parameters and one could re-tune them for every ensemble. In
our experience from different ensembles of SYM with gauge group SU(2) and
SU(3), the dependency on these parameters is very mild, and it is not
necessary to always find the optimal values. We have therefore kept fixed
the values stated here.

For the tuning of the smearing levels we have used SU(2) gauge ensembles at
$\beta=1.75$ and $\kappa=0.14925$. There are in total of 6800 thermalised
configurations of which we have measured every 64th for our tests. We use 40
stochastic estimators for the disconnected pieces.
 
\subsection{Presmearing of the gauge fields}  
\label{sec:preSmearing} 

\begin{figure}
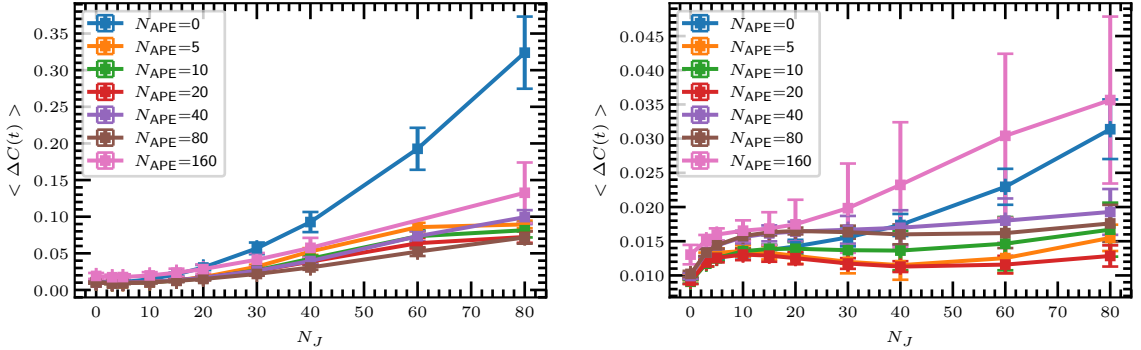

\input{figures/preSmeared_A-f0_disc.pgf}
\input{figures/preSmeared_A-f0.pgf}
\caption{Jackknife-error of the disconnected piece (left) and of the full
a-$f_{0}$ correlator (right) averaged over the interval $t \in [3,12]$
plotted against the Jacobi smearing level using different smearing levels
for the gauge-field $\tilde U$ in the smearing kernel
(\ref{eq:SmearingKernel}). Higher Jacobi smearing levels are affected by
larger errors. Using smeared gauge fields in the Jacobi smearing suppresses
this error significantly. The disconnected piece and the full correlator
have been normalised to 1 at $t=t_0$.}
\label{fig:PreSmearA-f0}
\end{figure}

\begin{figure}
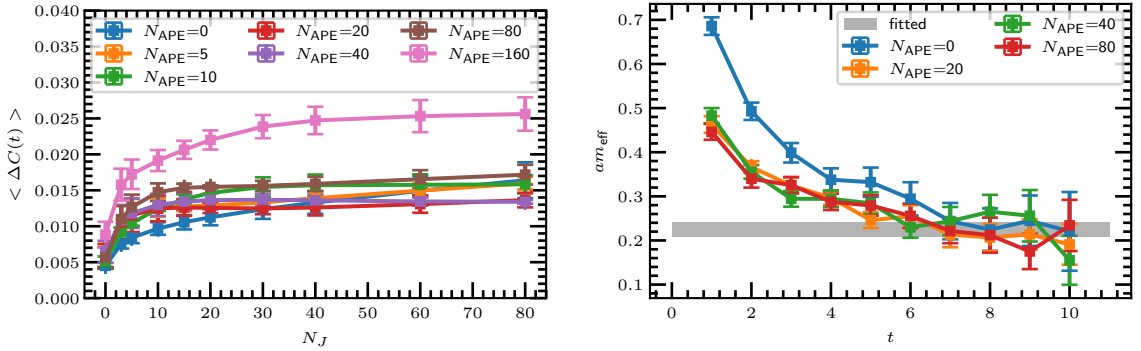

\input{figures/preSmeared_A-eta.pgf}     
\input{figures/effectivemass_A_eta_N_Ape.pgf}    
\caption{Left: Jackknife-error of the disconnected piece of the $\aetap$
correlator averaged over the interval $t \in [3,12]$ plotted against the
Jacobi smearing level, using different smearing levels for the gauge-field 
$\tilde U$ in the smearing kernel (\ref{eq:SmearingKernel}).
Right: Effective mass of the $\aetap$ correlator using different levels of APE
smeared gauge fields in the Jacobi smearing. The Jacobi smearing level is
fixed to $N_J=20$. Using smeared gauge fields more effectively suppresses
excited state contributions than Jacobi smearing with unsmeared gauge fields
(blue). The disconnected piece has been normalised to 1 at $t=t_0$.}
\label{fig:PreSmearEta}
\end{figure}

The application of Jacobi smearing defined with unsmeared gauge links
introduces additional noise to the signal of the disconnected contribution
to the a-$f_{0}$ meson, see Fig.~\ref{fig:PreSmearA-f0}. We have observed
that this noise can be suppressed by using smeared gauge fields in the
Jacobi smearing. We have tested APE and Stout smearing together with
different smearing levels for the preparation of the gauge field. Our
results indicate that $N_{\text{APE}}=20,~\epsilon_{\text{APE}}=0.5 $ leads 
to a noise reduction for Jacobi smearing levels up to $N_{J}=80$. 
The results are not very sensitive
to $N_{\text{APE}}$, and values of $N_{\text{APE}}$ between $10$ and $80$
all feature sufficient noise suppression. Larger values for $N_{\text{APE}}$
lead again to a degradation of the signal, see $N_{\text{APE}}=160$ in
Figs.~\ref{fig:PreSmearA-f0}, \ref{fig:PreSmearEta}. Using Stout smearing
instead of APE smearing didn't improve the noise suppression in our tests
and we therefore stay with APE smearing.

In the case of the $\aetap$ correlator, using smeared gauge fields in the
Jacobi smearing has a different effect than on the a-$f_{0}$ correlator. It
does not suppress the noise that is introduced by Jacobi smearing (see
Fig.~\ref{fig:PreSmearEta}). However, when smeared gauge fields are used
instead of unsmeared ones, Jacobi smearing more effectively suppresses
excited state contributions to the correlator. Again, the results are not
very sensitive to the exact value of $N_\text{APE}$ as long as it is in a
suitable range $ 10\leq N_\text{APE}\leq 80 $.

Considering both $\afn$  and $\aetap$ correlators, the smearing level
$N_{\text{APE}}=20$ is chosen, as it appears to be a good choice for
suppressing noise and excited state contributions.

\subsection{Variational operator basis}
\label{subsec:VarBasis}

A variational basis for the GEVP can be built from different smearing levels
of the interpolating operators. There is a trade-off between the cost
required to build a large variational basis from many different smearing
levels and the gain of new information from such operators. Each new
smearing level requires additional inversions for the gluino-glue correlator
and for the connected piece of the meson correlator. Therefore it is
important to find those smearing levels which are most relevant for the
extraction of the masses. For this purpose we consider the set of operators
constructed from the smearing levels $N_\text{Smearing} \in
\{0,5,15,..,95\}$, where Jacobi smearing is used for the meson interpolators
and APE smearing is used for the glueball and gluino-glue interpolators.
From these sets we have systematically chosen different subsets to analyse
which smearing levels allow the most efficient estimation of the lowest two
states in each sector.

For each number $n$ of operators we have picked the following sets:
\begin{enumerate}
\item
Small smearing levels: only the smallest $n$ smearing levels of the
full set are taken into account.
\item 
Large smearing levels: only the largest $n$ smearing levels of the full set
are taken into account.
\item 
Uniformly distributed: out of the full set we chose the smallest and the
largest one and $n-2$ additional smearing levels uniformly distributed
in between.
\item 
Mid-high, uniformly distributed: out of the full set we chose a medium level
(\mbox{$N_\text{Smearing}=35$}) and the largest one, and $n-2$ additional
smearing levels uniformly distributed between them.
\end{enumerate}

\begin{figure}
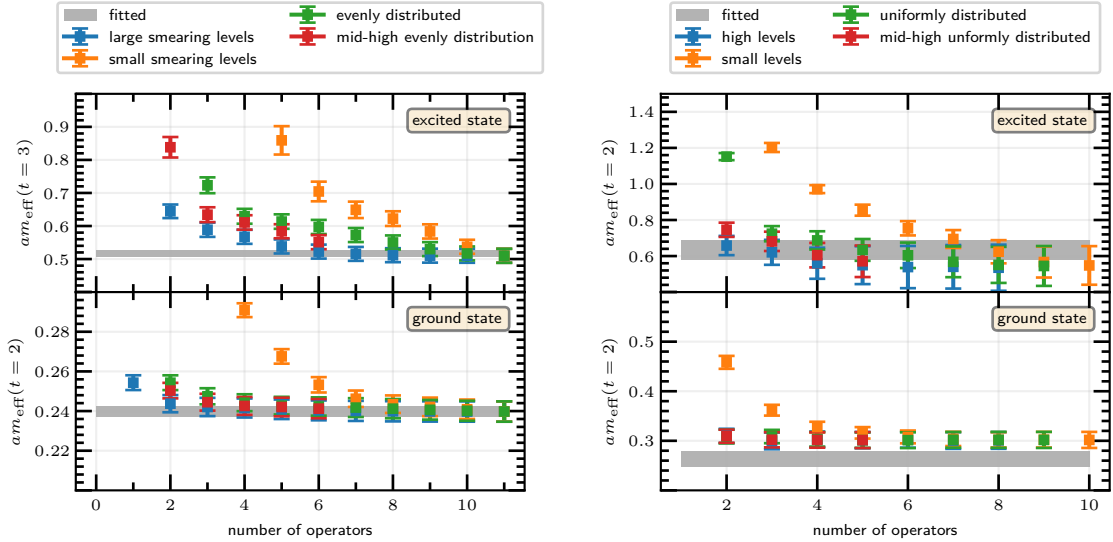

\input{figures/gluingoglue_smearing_levels_both.pgf} 
\input{figures/eta_smearinglevels_both.pgf} 
\caption{Effective masses of the lowest and first excited state of the
gluino-glue (left) and the $\aetap$ from the GEVP based on correlation
matrices of different sizes. Results for the matrices from the least-smeared 
operators are shown in orange, results from the most-smeared operators are shown in blue. The final result of the fit to the
eigenvalues is shown for comparison.}
\label{fig:gg_NAPEComp}
\end{figure}

To judge the quality of a mass determination with a chosen correlation
matrix, the effective masses of the lowest two states at fixed $t$ are used
as estimators. The results from the described procedure are shown in
Figs.~\ref{fig:gg_NAPEComp} and \ref{fig:enlarged_basis} together with the
best estimate for the mass obtained by fitting the eigenvalues using the
full set of interpolating operators.

The results for the dependency of the effective masses on the smearing
levels is similar for all three particles: smallest smearing levels
obviously lead to unwanted contributions of higher excitations so that the
effective masses at small $t$ are significantly higher than the best
estimate. Using only the highest smearing levels leads to the strongest
suppression of these excited state contributions. The uniformly distributed
subsets also suppress excited state contributions, but not quite as much as
the large smearing levels. Note that the error of the effective masses is
almost constant for the different combinations of the smearing levels. We
conclude that it is optimal to use a set of rather large smearing levels for
the mass estimations.

\subsection{Number of stochastic estimators} 
\label{sec:SETOptimization}

\begin{figure}
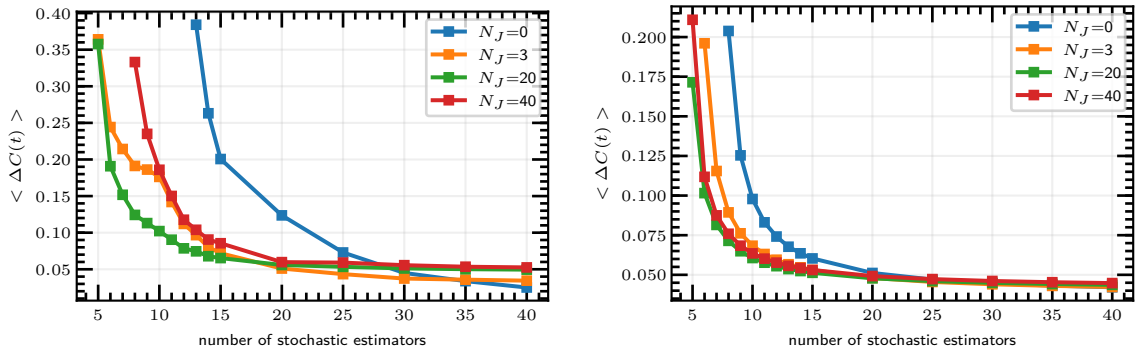

\input{figures/numEstimators_A-f0.pgf}  
\input{figures/numEstimators_A-eta.pgf}
\caption{Jackknife error estimates of the disconnected piece at averaged over the interval \mbox{$t \in [3,10]$} for
the scalar (left) and the pseudoscalar (right) meson as a function of the
number of stochastic estimators. For other values of $t$ the results are
similar. The disconnected pieces are normalised to $1$ at $t=t_0$.}
\label{fig:SET_estimators}
\end{figure}

With the optimised Jacobi smearing parameters we also tested the influence
of the number of SET estimators on the error of the disconnected pieces. For
this estimation we used every fourth configuration of our test ensemble. The
aim is to find a value for the number of stochastic estimators where the
error from the stochastic estimators is much smaller than the gauge noise.
We find that higher smearing levels require less stochastic estimators, see
Fig.~\ref{fig:SET_estimators}, and the scalar correlator requires more
stochastic estimators than the pseudoscalar one. At around 20 stochastic
estimators the error is dominated by the gauge noise (the stochastic error
is smaller than $15\%$ of the gauge noise) for all tested smearing levels
except for the unsmeared scalar correlator.
  
\section{Extended variational basis}
\label{sec:operatorbasis}

In general, an operator transforming according to a given irreducible
representation of the lattice symmetry group has a non-zero overlap with all
the eigenstates of the Hamiltonian with same quantum numbers. Mixing occurs
if two operators share the same transformation properties, independently of
their fermion or gluon field-content. In the case of the $0^{++}$ or
$0^{-+}$ channels the relevant operators can be constructed from
glueball-like combinations of Wilson loops and meson-like operators of the
form $\bar{\lambda}\Gamma\lambda$. Therefore the variational basis
(\ref{eq:GEVP}) can be enlarged to include the most general mixing between
glueball and meson operators.

Taking both kinds of operators into account, the full correlation matrix has
the following form
\begin{align} 
\label{eq:enlargedMatrix}
C(t)= & \left(
\begin{array}{cc}
\left\langle O_{\text{gb}}(t) O_{\text{gb}}^{\dagger}(0) \right\rangle  
 & \left\langle O_{\text{gb}}(t) O_{\text{meson}}^{\dagger}(0)\right\rangle \\
\left\langle O_{\text{meson}}(t) O_{\text{gb}}^{\dagger}(0) \right\rangle  
 & \left\langle O_{\text{meson}}(t) O_{\text{meson}}^{\dagger}(0)\right\rangle 
\end{array}\right).
\end{align}
Each entry of $C(t)$ is a submatrix consisting of the correlators among
different interpolating fields, where $O_{\text{gb}}$ stands for
glueball-like operators and $O_{\text{meson}}$ stands for meson-like
operators. Since the meson and the glueball operators are constructed from
quite different components, it is expected that their mutual overlap is
small. Using this larger correlation matrix we expect to obtain
significantly improved signals.

\begin{figure}
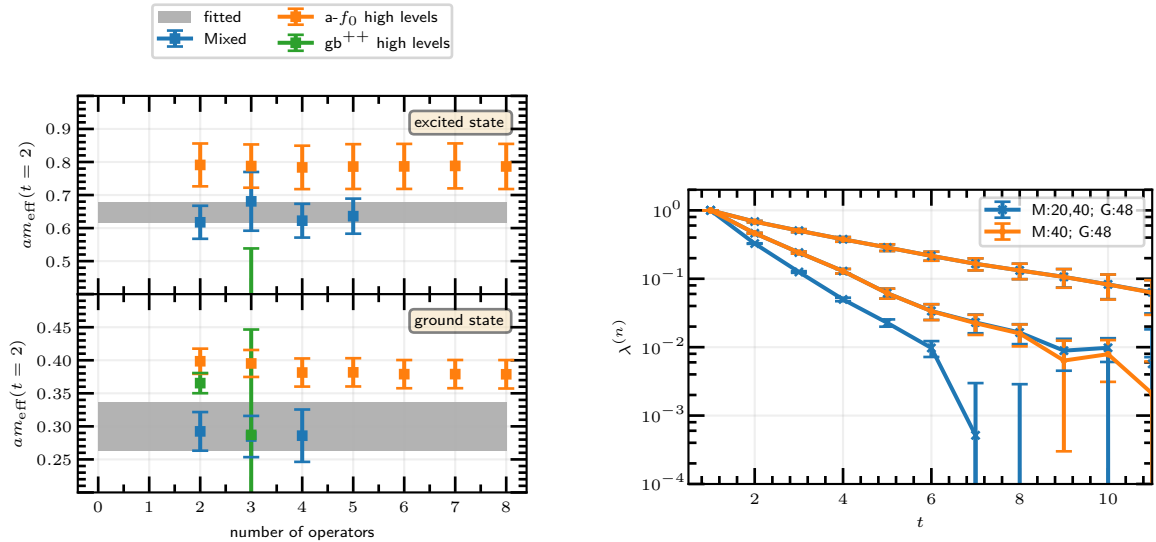

\input{figures/0++_smearing_levels_both.pgf}
\input{figures/excited_eigenvalues.pgf}  
\caption{Left: The ground state and the first excited state of the
$0^{++}$-channel are accessible with the mixed basis of one meson and one
glueball operator (blue). Adding further meson or glueball operators (blue) 
doesn't improve the projection to the state. If only meson or only
glueball operators are used (orange and green) excited contributions are not
as effectively suppressed as compared to the mixed basis.\protect \\ Right:
Shown are the first three generalised eigenvalues $\lambda^{(n)}$ of the
GEVP in the $0^{++}$-channel. In orange the results from using a mixed basis
including one meson operator (smearing level 40) and one glueball operator
(smearing level 48) are shown. Adding a further operator (mesonic, smearing
level 20) to the mixed basis (blue) doesn't change the signal for the ground
state and the first excited state, but a signal for a second excited state
appears.}
\label{fig:enlarged_basis}
\end{figure}

\textbf{Scalar $\mathbf{0^{++}}$ channel}: The results of our calculations
show that the enlarged variational basis leads to a more effective
suppression of excited states contributions than using meson or glueball
operators alone. Even the minimal choice of using only one meson and one
glueball operator appears to be sufficient to extract the masses of the
lowest two states in the scalar channel, see Fig.~\ref{fig:enlarged_basis}.
Using more than one glueball or more than one meson operator does not
improve this estimation, but provides better access to the excited states.
We therefore conclude that it is crucial to include both meson and glueball
operators in the variational basis to reliably extract the ground state in
the scalar channel.

\begin{figure}[t]
\begin{center}
\input{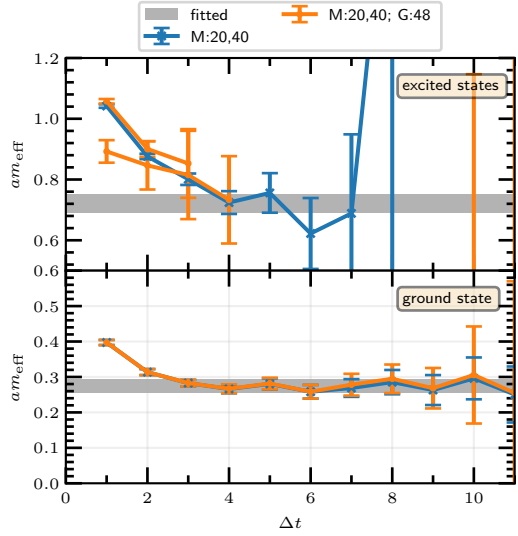}
\caption{Effective masses in the $0^{-+}$-channel from a basis consisting of
two mesonic operators (smearing levels 20 and 40) (blue), and a basis
including an additional glueball operator (smearing level 48) (orange). The
estimation of the ground state (lower panel) doesn't change when the
glueball operator is included. In the upper panel the first and second
excited states of the mixed basis are shown (orange). They are of similar
mass. Apparently one is the excited meson state (it aligns well with the
excited state of the purely mesonic basis (blue)), and the other one is the
glueball ground state.}
\label{fig:0-+}
\end{center} 
\end{figure}

\textbf{Pseudoscalar $\mathbf{0^{-+}}$channel}: In this channel the
estimation of masses does not improve when the enlarged basis also includes
glueball operators. The lowest two states can be analysed sufficiently by
using only meson operators, see Fig.~\ref{fig:0-+}, which means that the
glueball operators do not mix significantly with these states. This is in
accord with the observation that the entries in the off-diagonal blocks of
the correlation matrix (\ref{eq:enlargedMatrix}) are zero (within rather
large errors).
 
\section{Supermultiplets}
\label{sec:multiplet}

\begin{figure}
\begin{center}
\input{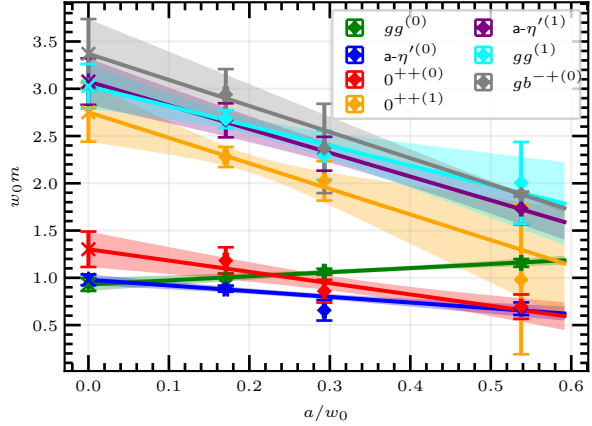}
\caption{Extrapolation to the continuum of the masses of the ground states 
and first excited states in the $0^{++}$, $0^{-+}$ and $1/2^{+}$ gluino-glue 
channels as explained in the text.}
\label{fig:Spectrum}
\end{center} 
\end{figure}

Using the methods explained above we could not only improve the precision of
our previous estimation of the ground state masses \cite{Bergner:2015adz},
but it also allowed to extract the masses of the first excited states, see
Fig.~\ref{fig:Spectrum}. Note, that in contrast to the analysis above, we
used a more conservative choice of smearing levels for the meson operators
only up to $N_J=40$. As explained above, including the glueball operator in
the GEVP for the pseudoscalar channel does not improve the results,
therefore we have analysed the pseudoscalar meson $\aetap$ and the glueball
$\text{gb}^{-+}$ separately. The results for the masses are collected in
Tab.~\ref{tab:masses}. The ground state masses in the $0^{-+}$ and $1/2^{+}$
channels, extrapolated to the continuum limit, agree quite well with each
other, while the one in the $0^{++}$ channel is somewhat higher, deviating
by $1.7\sigma$ from the mean value $w_0 m^{(0)}=0.98(5)$ of the
supermultiplet. Similarly to the ground states, the respective excited
states also seem to form a mass degenerate supermultiplet. Interestingly the
$0^{-+}$-glueball ground state and the excited state of the $\aetap$ both
have similar masses, but they appear as two independent states in the
variational method which do not mix.

\begin{table}
\begin{center}
\caption{Ground and excited state masses in lattice units extrapolated to
the chiral limit ($\kappa = \kappa_c$), and the extrapolations to the
continuum limit in units of the scale $w_0$. The subscript $0^{++}$ denotes
the result in the mixed channel, gg the gluino-glue, $\aetap$ the
pseudoscalar meson and $\text{gb}^{-+}$ the pseudoscalar glueball. The
masses in the continuum limit are given as $w_{0} m$.}
\begin{tabular}{|c|c|c|c|c|c|c|c|c|}
\hline 
$\beta$ & $w_{0}/a$ & $am^{(0)}_{0^{++}}$ & $am^{(0)}_{\aetap}$ 
   & $am^{(0)}_{\text{gg}}$ & $am^{(1)}_{0^{++}}$ & $am^{(1)}_{\aetap}$ 
   & $am^{(1)}_{\text{gg}}$ & $am^{(0)}_{\text{gb}^{-+}}$
\tabularnewline
\hline 
\hline 
1.6 & 1.860(4)  & 0.37(7) & 0.36(4) & 0.62(2) & 0.5(4) & 0.93(9) & 1.1(2) & 1.01(1)
\tabularnewline
\hline 
1.75 & 3.411(2)  & 0.25(4) & 0.19(3) & 0.312(9) & 0.59(6) & 0.67(5)  & 0.67(8) & 0.7(1) 
\tabularnewline
\hline 
1.9 & 5.858(8) & 0.20(2) & 0.151(5) & 0.17(1) & 0.38(2) & 0.45(3) & 0.46(1) & 0.50(5)
\tabularnewline
\hline 
\hline 
cont.   &  & 1.3(2) & 0.98(6) & 0.93(6) & 2.8(3) & 3.1(2)  & 3.0(2) & 3.4(4) 
\tabularnewline
\hline 
\end{tabular}
\label{tab:masses}
\end{center}
\end{table} 

\section{Mixing between Glueballs and Mesons}
\label{sec:mixing}

The numerical results of the extended variational approach indicate mixing
between glueball and mesonic states. Such a mixing has been predicted in the
literature on the structure of the lowest chiral supermultiplets
\cite{Farrar:1997fn,Farrar:1998rm}. From the calculated correlation matrices
we can get insights about the nature of the physical states with respect to
their mesonic and glueball content.

From the eigenvectors $\vec{v}_n$ of the GEVP (\ref{eq:GEVP}) the
corresponding physical states $\ket{n}$ can be reconstructed and decomposed
into a glueball contribution $\ket{\phi^{(g)}}$ and a meson contribution
$\ket{\phi^{(m)}}$:
\begin{align} 
\ket{n}&= \sum_{i=1}^{n_g} v_{ni}^{(g)} \hat O_i^{(g)}\ket{\Omega}
+\sum_{i=1}^{n_m} v_{ni}^{(m)} \hat O_i^{(m)}\ket{\Omega}\\
&= \sum_{i=1}^{n_g} v_{ni}^{(g)} \ket{\phi_i^{(g)}}
+ \sum_{i=1}^{n_m} v_{ni}^{(m)} \ket{\phi_i^{(m)}}
\doteq \ket{\phi_n^{(g)}}+\ket{\phi_n^{(m)}},
\end{align}
where $v^{(g)}_{ni}$ and $v^{(m)}_{ni}$ are the components of the
eigenvectors $\vec{v}_n$ corresponding to the glueball operators $O_i^{(g)}$
and the meson operators $O_i^{(m)}$, respectively, and $\ket{\Omega}$ is the
vacuum state. Note that $\ket{n},\ket{\phi^{(g)}}$ and $\ket{\phi^{(m)}}$
are not normalised here.

The inner products $c_{ni}\doteq \braket{\phi_i|n}$ can be calculated as the
vectors dual to the $\vec{v}_n$ by means of~\cite{Blossier:2009kd}
\begin{equation}
\sum_i v^*_{mi} c_{ni}=\delta_{mn}.
\end{equation}
So they are the row vectors of $M^{-1}$, where $M$ is the matrix formed by
the column vectors $\vec{v}_n$. Let us denote $c_{ni}^{(g)} =
\braket{\phi_i^{(g)}|n}$ and $c_{ni}^{(m)} = \braket{\phi_i^{(m)}|n}$ the
restrictions of the $c_{ni}$ to the glueball and the meson components,
respectively. The normalisations of the vectors can be obtained as
\begin{align}
{N_{n}^{(g)}}^2 &=\braket{\phi_n^{(g)}|\phi_n^{(g)}}
=\sum_{ij} v^{*(g)}_{ni} v_{nj}^{(g)}\braket{\phi_i^{(g)}|\phi_j^{(g)}}
=\sum_{ij} v^{*(g)}_{ni} v_{nj}^{(g)} \sum_k c^{(g)}_{ki} c_{kj}^{*(g)}\\  
{N_{n}^{(m)}}^2 &=\braket{\phi_n^{(m)}|\phi_n^{(m)}}
=\sum_{ij} v^{*(m)}_{ni} v_{nj}^{(m)}\sum_k c^{(m)}_{ki} c_{kj}^{*(m)}\\
N_n^2 &=\braket{n|n}
=\sum_{ij} v^{*}_{ni} v_{nj} \sum_k c_{ki} c_{kj}^{*} \,.
\end{align}
Now we define the glueball and the meson contents of the physical state
$\ket{n}$ as the overlap of this state with the glueball and meson
contributions
\begin{align}
c_{n}^{(g)}&\doteq \frac{1}{N^{(g)}_n N_n}\braket{\phi^{(g)}|n}
=\frac{1}{N^{(g)}_n N_n}\sum_i v_{ni}^{*(g)}c^{(g)}_{ni}\\
c_{n}^{(m)}&\doteq \frac{1}{N^{(m)}_n N_n}\braket{\phi^{(m)}|n}
=\frac{1}{N^{(m)}_n N_n}\sum_i v_{ni}^{*(m)}c^{(m)}_{ni} .
\end{align}
We would like to point out that the glueball contribution $\ket{\phi^{(g)}}$
and the meson contribution $\ket{\phi^{(m)}}$ are not necessarily orthogonal
to each other and therefore $c_{n}^{(g)2}$ and $c_{n}^{(m)2}$ in
general do not add up to 1.

\begin{figure}
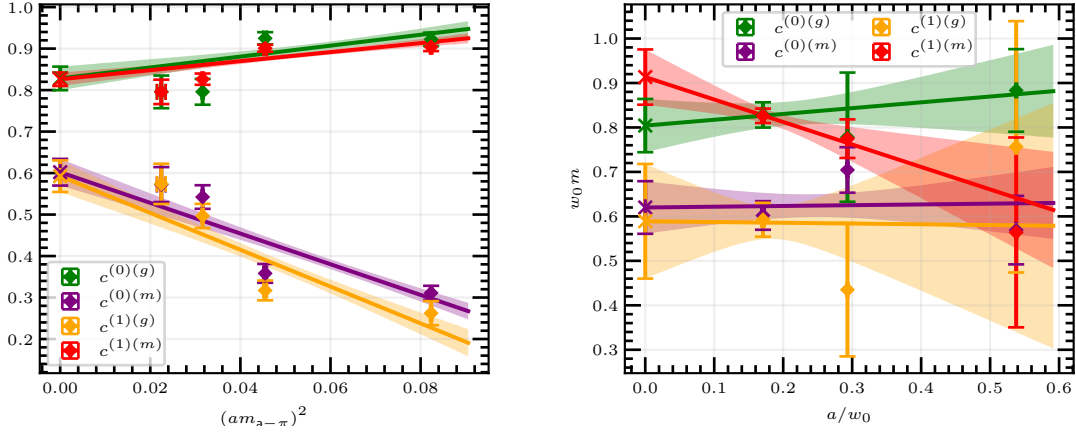

\input{figures/chiralExtrMixings_a_1_9.pgf}
\input{figures/continuumExtrMixings_w0_1.pgf}
\caption{Left: chiral extrapolation of the glueball and meson contents
$c^{(g)}$ and $c^{(m)}$ in the $0^{++}$ ground state (0) and excited state
(1) for our finest lattices ($\beta=1.9$). Right: Extrapolation of
the overlaps at different lattice spacings to the continuum limit.}.
\label{fig:0++-mixing}
\end{figure}

We have determined the glueball and meson contents $c_{n}^{(g)}$ and
$c_{n}^{(m)}$ of the ground state and excited state in the $0^{++}$ channel,
see Fig.~\ref{fig:0++-mixing} and Tab.~\ref{tab:mixing}. 
Extrapolated to the continuum limit, we find
that the ground state has a glueball content of $c^{(g)}=0.80(6)$ and a
meson content of $c^{(m)}=0.62(6)$, meaning that the ground state in the
scalar channel shows significant mixing of glueball and meson contents. The
excited state appears to have the opposite glueball and meson contents than
the ground state, namely $c^{(g)}=0.6(1)$ and $c^{(m)}=0.91(6)$. While the
ground state is more glueball like, the excited state is more meson like.
This is especially visible in the chiral extrapolation at our smallest
lattice spacing, Fig.~\ref{fig:0++-mixing}. Interestingly the squares
$c^{(g)2}$ and $c^{(m)2}$ do add up to 1 within errors as in the
orthogonal case.

\begin{table}
\begin{center}
\caption{Meson and glueball contents of the ground state~(0) and excited
state(1) in the scalar channel, extrapolated to the chiral limit.}
\begin{tabular}{|c|c|c|c|c|}
\hline 
$\beta$ & $c^{(0)(m)}$ & $c^{(0)(g)}$ & $c^{(1)(m)}$ & $c^{(1)(g)}$
\tabularnewline
\hline 
\hline 
1.6 & 0.57(7) & 0.88(9) & 0.6(2) & 0.8(3)\tabularnewline
\hline 
1.75 & 0.70(5) & 0.8(1)  & 0.77(4)  & 0.4(2)\tabularnewline
\hline 
1.9 & 0.62(5)  & 0.80(6) & 0.82(2)  & 0.59(4)\tabularnewline
\hline 
continuum & 0.63(3) & 0.82(3) & 0.91(6)  & 0.6(1) \tabularnewline
\hline 
\end{tabular}
\label{tab:mixing}
\end{center}
\end{table}

\section{Conclusions}

In this work we have presented in detail our improved techniques for the
estimations of bound state masses in $\mathcal{N}=1$ supersymmetric
Yang-Mills theory. We have enlarged the basis of operators in our
variational analysis, combining different glueball and mesonic operators.
The enlarged basis allows for the first time an investigation of the mixing between these
two classes of operators. The new techniques have improved our estimates of
the ground state masses and allowed an estimation of the first excited
states, which have not been accessed in previous studies. We have combined
the results for the masses of the ground states and first excited states
from several different lattice spacings in an extrapolation to the continuum
limit. The first interesting observation is the formation of
supermultiplets. The ground state masses of the scalar, pseudoscalar, and
fermion channel become approximately degenerate when extrapolated to the
chiral and continuum limit. This is in line with our previous results and
indicates the formation of a chiral supermultiplet of lightest states. The
deviation of the scalar mass from the average value is a bit larger than the
quoted error, but this is most likely due to underestimated systematic
uncertainties, since only at a single lattice spacing there is a significant
deviation between the scalar and pseudoscalar channel. The average mass of
the lightest supermultiplet is around $w_0 m^{(0)}=0.98(5)$.

Another very interesting result is the possible formation of a chiral
supermultiplet of excited states at $w_0 m^{(1)} \approx 3.1$. The masses of
these states in lattice units are around $0.5\, a^{-1}$, i.\,e.\ around half
of the cutoff scale. It is quite unexpected that states at these high
energies are not more affected by the supersymmetry breaking lattice
artefacts. For a complete analysis a more detailed investigation of the
bound state spectrum is required, including also other multiplets and a
larger set of quantum numbers. We are currently adding, for instance, an
investigation of the baryonic states of the theory~\cite{Ali:2018rln}.

We have investigated the mixing between glueball and mesonic operators in
the scalar and pseudoscalar channels. In the scalar channel we have found
significant mixing between glueball and meson contents. In the pseudoscalar
channel we found no hints of a significant mixing. The pseudoscalar ground
state is clearly dominated by the mesonic contribution, whereas the scalar
ground state has an apparent predominant glueball contribution. 

Our results might help for a better understanding of the low energy 
effective action~\cite{Veneziano:1982ah,Farrar:1997fn,Farrar:1998rm} and the
conjectures presented in~\cite{Evans:1997jy,Feo:2004mr}.
In~\cite{Farrar:1997fn,Farrar:1998rm}, which refines and extends the
analysis of~\cite{Evans:1997jy}, the nature of the lowest two chiral
supermultiplets has been investigated on the basis of an effective action
including mesonic and glueball-type degrees of freedom. The lightest states
were conjectured to be of the glueball type, if mixing is not too strong.
The argument is based on the perturbation of the effective theory by a small
gluino mass, which leads to a splitting of the multiplets.
In the mesonic multiplet the pseudoscalar meson becomes lighter than the 
scalar meson. Drawing on the proof~\cite{West:1995ym} that the lightest scalar 
state, which has overlap with the $0^{++}$ glueball operator, is not heavier 
than the lightest pseudoscalar state, which has overlap with the $0^{-+}$ 
glueball operator, they conclude that the multiplet of glueball states must be 
lighter than the multiplet of mesons. 

On the other hand, in Ref.~\cite{Feo:2004mr} the lightest states are conjectured 
to be of mesonic type with a small mixing of the glueballs. Their analysis 
employs an effective Lagrangian with an arbitrary mixing angle between the 
glueball ($R$-charge 0) and meson ($R$-charge 2) multiplet. 
Generally, the effective Lagrangian allows either mesonic or glueball-type
states to be the lightest ones, depending on an unknown parameter.
The argument for the ordering of states is then based on the large-$N_c$
equivalence of SU($N_c$) SYM and QCD-like theories. In QCD the mesonic
$\eta'$ appears to be much lighter than the scalar glueball, which leads to
the conjectures about the ordering of states and only a small mixing for
SYM. A distinction between the scalar and pseudoscalar channels is, however,
not being made.

Our numerical findings in the scalar channel, where the lightest state is
dominantly of glueball type, are consistent with the predictions of 
\cite{Farrar:1997fn,Farrar:1998rm,Evans:1997jy}. On the other hand,
it appears that the pseudoscalar ground state is dominated by the mesonic 
contribution with a negligible mixing of the glueball, which would correspond
to the scenario advocated in~\cite{Feo:2004mr}. Also, in contrast to the
above arguments, we do not find the same mixing angle for the scalar and the 
pseudoscalar channel. A detailed consideration of the effects of a small
gluino mass could shed more light on these questions.

\section{Acknowledgments}

The authors gratefully acknowledge the Gauss Centre for Supercomputing
e.~V.\,\linebreak(www.gauss-centre.eu) for funding this project by providing
computing time on the GCS Supercomputers JUQUEEN and JURECA at J\"ulich Supercomputing
Centre (JSC) and SuperMUC at Leibniz Supercomputing Centre (LRZ). Further
computing time has been provided the compute cluster PALMA of the University of M\"unster. This work is
supported by the Deutsche Forschungsgemeinschaft (DFG) through the Research
Training Group ``GRK 2149: Strong and Weak Interactions - from Hadrons to
Dark Matter''. G.~Bergner acknowledges support from the Deutsche
Forschungsgemeinschaft (DFG) Grant No.\ BE 5942/2-1. S.~Ali acknowledges
financial support from the Deutsche Akademische Austauschdienst (DAAD).
P.~Scior acknowledges support by the Deutsche Forschungsgemeinschaft (DFG)
through the CRC-TR 211 ``Strong-interaction matter under extreme
conditions''– project number 315477589 – TRR 211.


\begin{thebibliography}{99}

\bibitem{Veneziano:1982ah}
G.~Veneziano and S.~Yankielowicz,
\emph{An effective Lagrangian for the pure $\mathcal{N}=1$ supersymmetric
Yang-Mills theory},
Phys.\ Lett.\ B \textbf{113} (1982) 231.

\bibitem{Farrar:1997fn}
G.~R.~Farrar, G.~Gabadadze and M.~Schwetz,
\emph{On the effective action of N=1 supersymmetric Yang-Mills theory},
Phys.\ Rev.\ D \textbf{58} (1998) 015009
[\arxiv{hep-th/9711166}\,].

\bibitem{Farrar:1998rm}
G.~R.~Farrar, G.~Gabadadze and M.~Schwetz,
\emph{The spectrum of softly broken N=1 supersymmetric Yang-Mills theory},
Phys.\ Rev.\ D \textbf{60} (1999) 035002
[\arxiv{hep-th/9806204}\,].

\bibitem{Demmouche:2010sf}
K.~Demmouche, F.~Farchioni, A.~Ferling, I.~Montvay, G.~M\"unster, E.~E.~Scholz 
and J.~Wuilloud,
\emph{Simulation of 4d N=1 supersymmetric Yang-Mills theory with Symanzik improved 
gauge action and stout smearing},
Eur.\ Phys.\ J.\ C \textbf{69} (2010) 147
[\arxiv{1003.2073} [\texttt{hep-lat}]].

\bibitem{Bergner:2015adz}
G.~Bergner, P.~Giudice, I.~Montvay, G.~M\"unster and S.~Piemonte,
\emph{The light bound states of supersymmetric SU(2) Yang-Mills theory},
JHEP \textbf{1603} (2016) 080 
[\arxiv{1512.07014} [\texttt{hep-lat}]].

\bibitem{Bergner:2013nwa}
G.~Bergner, I.~Montvay, G.~M\"unster, U.~D.~\"Ozugurel and D.~Sandbrink,
\emph{Towards the spectrum of low-lying particles in supersymmetric Yang-Mills 
theory}, 
JHEP \textbf{1311} (2013) 061
[\arxiv{1304.2168} [\texttt{hep-lat}]].

\bibitem{Bergner:2012rv} 
G.~Bergner, T.~Berheide, I.~Montvay, G.~M\"unster, U.~D.~\"Ozugurel and
D.~Sandbrink,
\emph{The gluino-glue particle and finite size effects in supersymmetric Yang-Mills 
theory},
JHEP \textbf{1209} (2012) 108
[\arxiv{1206.2341} [\texttt{hep-lat}]].

\bibitem{Bergner:2014ska} 
G.~Bergner, P.~Giudice, I.~Montvay, G.~M\"unster and S.~Piemonte,
\emph{Influence of topology on the scale setting},
Eur.\ Phys.\ J.\ Plus \textbf{130} (2015) 229
[\arxiv{1411.6995} [\texttt{hep-lat}]].

\bibitem{Bergner:2015cqa}
G.~Bergner, P.~Giudice, G.~Münster and S.~Piemonte,
\emph{Witten index and phase diagram of compactified $\mathcal N=1$ supersymmetric 
Yang-Mills theory on the lattice},
PoS(LATTICE 2015) 239
[\arxiv{1510.05926} [\texttt{hep-lat}]].

\bibitem{Curci:1986sm}
G.~Curci and G.~Veneziano, 
\emph{Supersymmetry and the lattice: a reconciliation?},
Nucl.\ Phys.\ B \textbf{292} (1987) 555.

\bibitem{Ali:2018fbq} 
S.~Ali, H.~Gerber, I.~Montvay, G.~M\"unster, S.~Piemonte, P.~Scior and G.~Bergner,
\emph{Analysis of Ward identities in supersymmetric Yang–Mills theory},
Eur.\ Phys.\ J.\ C \textbf{78} (2018) 404
[\arxiv{1802.07067} [\texttt{hep-lat}]].

\bibitem{Munster:2014cja}
G.~M\"unster and H.~St\"uwe,
\emph{The mass of the adjoint pion in $\mathcal{N}=1$ supersymmetric Yang-Mills
theory},
JHEP \textbf{1405} (2014) 034
[\arxiv{1402.6616} [\texttt{hep-th}]].

\bibitem{Bergner:2017ytp}
G.~Bergner and S.~Piemonte,
\emph{The running coupling from gluon and ghost propagators in the Landau gauge: 
Yang-Mills theories with adjoint fermions}, 
Phys.\ Rev.\ D \textbf{97} (2018) 074510
[\arxiv{1709.07367} [\texttt{hep-lat}]].

\bibitem{Luscher:2010iy}
M.~Lüscher,
\emph{Properties and uses of the Wilson flow in lattice QCD},
JHEP \textbf{1008} (2010) 071,
Erratum: [JHEP \textbf{1403} (2014) 092]
[\arxiv{1006.4518} [\texttt{hep-lat}]].

\bibitem{Borsanyi:2012zs}
S.~Borsanyi \textit{et al.},
\emph{High-precision scale setting in lattice QCD},
JHEP \textbf{1209} (2012) 010
[\arxiv{1203.4469} [\texttt{hep-lat}]].

\bibitem{Montvay:2005tj}
I.~Montvay and E.~Scholz,
\emph{Updating algorithms with multi-step stochastic correction},
Phys.\ Lett.\ B \textbf{623} (2005) 73
[\arxiv{hep-lat/0506006}\,].

\bibitem{Bergner:2011zp}
G.~Bergner and J.~Wuilloud,
\emph{Acceleration of the Arnoldi method and real eigenvalues of the non-Hermitean 
Wilson-Dirac operator},
Comput.\ Phys.\ Commun.\ \textbf{183} (2012) 299
[\arxiv{1104.1363} [\texttt{hep-lat}]].

\bibitem{Luscher:1990ck}
M.~L\"uscher and U.~Wolff,
\emph{How to calculate the elastic scattering matrix in two-dimensional quantum 
field theories by numerical simulation}, 
Nucl.\ Phys.\ B \textbf{339} (1990) 222.

\bibitem{Blossier:2009kd}
B.~Blossier, M.~Della Morte, G.~von Hippel, T.~Mendes and R.~Sommer,
\emph{On the generalized eigenvalue method for energies and matrix elements in 
lattice field theory},
JHEP \textbf{0904} (2009) 094
[\arxiv{0902.1265} [\texttt{hep-lat}]].

\bibitem{Dong:1993pk}
S.~J.~Dong and K.~F.~Liu,
\emph{Stochastic estimation with Z(2) noise},
Phys.\ Lett.\ B \textbf{328} (1994) 130
[\arxiv{hep-lat/9308015}\,].

\bibitem{Bergner:2011zz} 
G.~Bergner, I.~Montvay, G.~M\"unster, U.~D.~\"Ozugurel and D.~Sandbrink,
\emph{Supersymmetric Yang-Mills theory: A step towards the continuum},
PoS(Lattice 2011) 055
[\arxiv{1111.3012} [\texttt{hep-lat}]].

\bibitem{Albanese:1987ds}
M.~Albanese \textit{et al.} [APE Collaboration],
\emph{Glueball masses and string tension in lattice QCD},
Phys.\ Lett.\ B \textbf{192} (1987) 163.
  
\bibitem{Gusken:1989ad}
S.~G\"usken, U.~L\"ow, K.~H.~M\"utter, R.~Sommer, A.~Patel and K.~Schilling,
\emph{Non-singlet axial vector couplings of the baryon octet in lattice QCD},
Phys.\ Lett.\ B \textbf{227} (1989) 266.

\bibitem{Ali:2017iof} 
S.~Ali, G.~Bergner, H.~Gerber, P.~Giudice, S.~Kuberski, I.~Montvay,
G.~M\"unster and S.~Piemonte,
\emph{Supermultiplets in $\mathcal{N}=1$ SUSY SU(2) Yang-Mills theory}, 
EPJ Web Conf.\ \textbf{175} (2018) 08016
[\arxiv{arXiv:1710.07464} [\texttt{hep-lat}]].

\bibitem{Ali:2018rln}
S.~Ali, G.~Bergner, H.~Gerber, C.~Lopez, I.~Montvay, G.~M\"unster, S.~Piemonte 
and P.~Scior,
\emph{Baryonic states in supersymmetric Yang-Mills theory},
PoS(LATTICE2018) 207
[\arxiv{arXiv:1811.02297} [\texttt{hep-lat}]].

\bibitem{Evans:1997jy}
N.~J.~Evans, S.~D.~H.~Hsu and M.~Schwetz,
\emph{Lattice tests of supersymmetric Yang-Mills theory?},
[\arxiv{hep-th/9707260}\,].

\bibitem{Feo:2004mr}
A.~Feo, P.~Merlatti and F.~Sannino,
\emph{Information on the super Yang-Mills spectrum},
Phys.\ Rev.\ D \textbf{70} (2004) 096004
[\arxiv{hep-th/0408214}\,].

\bibitem{West:1995ym} 
G.~B.~West,
\emph{Theorem on the lightest glueball state},
Phys.\ Rev.\ Lett.\ \textbf{77} (1996) 2622
[\arxiv{hep-ph/9603316}\,].

\end{thebibliography}
\end{document}